\documentclass[aip,jcp,reprint,superscriptaddress,footinbib]{revtex4-1}
\usepackage{graphicx}
\usepackage{comment}
\usepackage{amsmath}
\usepackage{wasysym}
\usepackage[usenames, dvipsnames]{color}
\usepackage{braket}
\usepackage{hyperref}
\usepackage{epstopdf}
\usepackage{subcaption}
\usepackage[symbol]{footmisc}

\begin{document}


\title{Energy Transfer and Thermoelectricity in Molecular Junctions in Non-Equilibrated Solvents}

\author{Henning Kirchberg}
\affiliation{Department of Chemistry, University of Pennsylvania, Philadelphia, Pennsylvania 19104, United States of America}
\email{khenning@sas.upenn.edu}
\author{Abraham Nitzan}
\affiliation{Department of Chemistry, University of Pennsylvania, Philadelphia, Pennsylvania 19104, United States of America}
\email{anitzan@sas.upenn.edu}

\date{\today}

\begin{abstract}
We consider a molecular junction immersed in a solvent where the electron transfer is dominated by Marcus-type steps. However, the successive nature of the charge transfer through the junction does not imply that the solvent reach thermal equilibrium throughout the transport. In our previous work \cite{kir2020} we have determined the nonequilibrium distribution of the solvent where its dynamics, expressed by a friction, is considered in two limiting regimes of fast and slow solvent relaxation. In dependence of the nonequilibrium solvent dynamics, we investigate now the electrical, thermal and thermoelectric properties of the molecular junction. We show that by suitable tuning the friction, we can reduce the heat dissipation into the solvent and enhance the heat transfer between the electrodes. Interestingly, we find that the Seebeck coefficient grows significantly by adapting the solvent friction in both regimes. 
\end{abstract}


\maketitle

\section{Introduction}
Molecular electronics became a well established and quickly developing field in the last two decades. Presently, it provides a general platform to realize atomic-scale "heat engines" which convert electric energy into heat transfer, i.e., heating or cooling of adjoining reservoirs (Peltier effect), or where the flow of heat can be converted into usable electrical power (Seebeck effect) \citep{avi1974,nit2003,cor2007,ber2010,cui2018}. The basic block of such an engine, which we eventually address to, is a single-molecule junction liked to two metallic electrodes. The electron transfer through this junctions is driven by a voltage or thermal gradient across the electrodes. In addition, in transfer junctions immersed inside a (dielectric) solvent, the electron transport properties may be affected. The two extreme limits of this electron transport are, on one hand, elastic (tunneling and resonance) transfer in absence of appreciable interaction with the nuclear solvent
environment, which might be included as perturbation, and on the other, a sequence of hopping processes through one or more redox states where the electron is transiently localized by distorting its local environment. The interplay between electronic and (solvent) nuclear motions has been addressed in several studies investigating its impact on charge transport \cite{koc2005,har2011}. The coupling of electrons to the nuclear background can result in several interesting effects including Franck-Condon blockade, negative differential conductance, and rectification \cite{koc2005,har2011,gal2005}. The impact on heat generation and transport through molecular junction due to electron-nuclear interaction has been studied and analyzed both experimentally and theoretically \cite{gal2005,ren2012}. 

Most studies of electron-nuclear interaction in molecular junctions exploit the elastic electron transport as starting point and treat the interaction with the nuclear background as perturbation \cite{ren2012,koc2014,zim2014}. In contrast, molecular electron transfer processes are usually described by Marcus theory \cite{mar1956a,mar1956b,mar1986} and its extensions, in which the electron transfer kinetics is dominated nuclear motion between the initial and final nuclear configurations. Molecular conduction junctions that operate in the sequential hopping limit, where the transferring electron can be transiently localized on a redox site on the bridging molecule, are often also described within such pictures \cite{zha2008,mig2011,mig2012,kir2020}. In such cases, an important departure from Marcus type rate theory stems from the fact that transient electron localization on the bridge may be incomplete in the sense that subsequent hopping takes place before full environmental relaxation (polaron formation) is achieved. In such a case, electron transport through the molecular junctions depends on the rate on this relaxation while the relaxation itself has to be described self-consistently with the electron hopping process. We have recently discussed the consequences of this interplay between hopping and relaxation on the conduction and noise properties of a model molecular junction in which solvent relaxation is described by a Smoluchowsky process \cite{kir2020}. 
	
While Marcus electron transfer processes have been under study for nearly seventy years, their energetic aspects were rarely addressed. We have recently discussed several such aspects, showing that electron energy transfer between sites with different local temperature is accompanied by heat transfer \cite{cra2017}. In molecular junctions, elastic electron transmission is accompanied by the inter-electrodes heat transfer as well as electrical-heat energy conversion \cite{cui2017}. In the present of solvent relaxation such energy balance considerations must include energy exchange with the solvent which in turn affects thermal energy exchange processes at the electrodes. This problem has been recently addressed \cite{zim2020} within a model that assumes complete solvent relaxation during the intermediate localization on the bridge, namely describes conduction in molecular junctions as a sequence of Marcus processes. Here we generalize this calculation, taking the finite nature of the solvent relaxation rate into account. Electron and heat transport, as well as thermoelectric processes become in this case dependent of the solvent relaxation rate which we study in the small and large friction limits. This dependence opens the possibility of affecting the junction electric and thermoelectric properties  by controlling solvent friction.

This paper is organized as follows: First (Section II\& III) we review our recent work that describes our model and studies solvent effects on conduction in molecular junctions in the small and large friction limits. In Section IV we extend the Monte Carlo method introduced in Ref. \cite{kir2020} to the calculation of heat currents between the molecular state and the  surrounding baths – the two electrodes and the solvent. This method is used in Section V to evaluate the current-voltage characteristics of our junction (for both electronic and thermal currents) as function of solvent friction for in the large and small friction regimes. Finally, in Section VI we address the thermoelectric behavior of our model junction, specifically, in Section VII, showing results for the solvent-friction dependence of the Seebeck coefficient. Section VIII concludes.

\section{Model}
We consider a molecular junction where electrons move between two electrodes through a molecular bridge. We model the molecule as a two-state system that corresponds to their electronic occupation. For specificity we refer to these states an "unoccupied" $A$ and "occupied" $B$ state that contains $N-1$ and $N$ electrons, respectively. Limiting our consideration to two such states amount to the (standard) assumption that strong Coulomb interaction put states with other electronic occupation in energy regimes that are not accessible under the experimental conditions \cite{zha2008}. The molecule is furthermore embedded in a polar solvent which which dynamically responds to the different charging state of the molecule. The polar solvent continuously fluctuates due to the translational and reorientational motions of its
constituent molecules and, so, the polarization field.
In the Levich \cite{lev1959,lev1961} version of electron transfer theory, this physical model is represented by
a standard spin-boson model, with the solvent represented as bosonic environment coupled linearly to the molecular electronic occupation. Marcus further developed this theory \cite{mar1956a,mar1956b} by expressing the fluctuating polarization field by a distribution of solvent configurations along a single reaction coordinate, x, determined by harmonic free energy surfaces that depend on the molecular electronic state according to\cite{NitzanBook}

\begin{align}
E_A(x,\epsilon)&=E_A+\frac{1}{2} \hbar  \omega_0 x^2+\epsilon , \label{eq2}\\ 
E_B(x)&=E_B+\frac{1}{2} \hbar \omega_0(x-d)^2.
\label{eq3}
\end{align}
In this shifted harmonic surfaces model, $E_A$ and $E_B$ are the electronic energies at the equilibrium solvent configurations, chosen as $x_A=0$ and $x_B=d$ for the state $A$ and $B$ respectively. 
The harmonic forms and the identical curvatures of these surfaces correspond to the assumption that the solvent responds linearly to the charging state of the molecule and has the consequence that the reorganization energy
\begin{align}\label{reorgen}
E_R=\frac{1}{2}\hbar \omega_0 d^2
\end{align}
is the same irrespective on the process direction from $A$ to $B$ or vice versa. 

Following Marcus, electron transfer is treated in the high temperature limit so that nuclear dynamics is described classically. The transfer dynamics corresponds to the non- adiabatic limit of the Landau-Zenner expression for the probability to interchange between molecular state $A$ and $B$ \cite{lan1932,zen1932}. Under these assumptions, electron transfer events are dominated by solvent configurations where $E_A(x)=E_B(x)$, namely at the transition point along the reaction coordinate given by
\begin{align}
x_{TR}=\frac{E_B-E_A-\epsilon + \frac{1}{2} \hbar \omega_0 d^2}{\hbar \omega_0 d}\, .
\label{eq4}
\end{align}

In the following, we further account for finite solvent relaxation between electron hopping events like in our recent work \cite{kir2020}. In this picture the free energy surfaces $E_A(x)$ and $E_B(x)$ are used as potential energy surfaces for the reaction or solvent coordinate $x$. Provided that account is taken for the fact that this coordinate can exchange energy with all other solvent degrees of freedom, finite solvent relaxation implies that the coordinate can not be represented by a thermal equilibrium or Boltzmann distribution when a new electron transfer process takes place.

\section{Theoretical approach}

In order to account for finite solvent relaxation, we describe the probability distribution for the position and velocity of the reaction coordinate $P_j(x,v;t)$  for the molecular state $j=A,B$, by the Fokker-Planck equation
\begin{align}
\label{eqFokker}
\frac{\partial P_j (x,v;t)}{\partial t}=&\omega_0 \frac{d \bar{V}_j}{d x}\frac{\partial P_j}{\partial v}-\omega_0 v \frac{\partial P_j}{\partial x} \\ \notag &+\gamma \bigg [ \frac{\partial}{\partial v}(vP_j)+\frac{k_B T}{\hbar \omega_0} \frac{\partial^2 P_j}{\partial v^2} \bigg ]\, .
\end{align}
In Eq.\ (\ref{eqFokker}), the normalized potential surfaces are $\bar{V}_j=V_j/(\hbar \omega_0)$ with $V_A(x)=\frac{1}{2}\hbar \omega_0 x^2$ and $V_B(x)=\frac{1}{2}\hbar \omega_0 (x-d)^2$. Note that the position and velocity variables in Eqs.\ (\ref{eq2}),(\ref{eq3}) and (\ref{eqFokker}) are dimensionless. The solvent properties that enter at this level of description are manifested via the parameters $\omega_0$ and $\gamma$ that can be obtained from fitting of the observed dielectric response of the solvent to standard dielectric response models \cite{MayBook}. The solvent-molecule coupling enters via the parameter $d$ that determines the solvent reorganization energy $E_R$ as given in Eq.\ (\ref{reorgen}). 

As in our recent work \cite{kir2020}, we consider the implications of this dynamics in two limits. In the overdamped limit, $\gamma \gg \omega_0$, Eq.\ (\ref{eqFokker}) leads to a Smoluchowski equation, that describes diffusion along the $x$ coordinate, 
\begin{align}
\label{eqFokker1}
\frac{\partial P_j (x,t)}{\partial t} =\frac{\omega_0}{\hbar \beta \gamma} \frac{\partial}{\partial x} \bigg [ \frac{\partial}{\partial x}+\beta \hbar \omega_0 \frac{d \bar{V}_j}{d x} \bigg] P_j(x,t)\, ,
\end{align}
where $\beta=(k_BT)^{-1}$. In the opposite underdamped limit, $\gamma \ll \omega_0$, the relaxation implied by Eq.\ (\ref{eqFokker}) may be reduced, after phase averaging, to diffusion in energy space, which is described by 
\begin{align}
\label{eqFokker2}
\frac{\partial P_j (E,t)}{\partial t} =\frac{\partial}{\partial E} \bigg [ \gamma E \bigg[1+k_B T \frac{\partial}{\partial E}\bigg] P_j(E,t)\bigg]\, .
\end{align}

The distribution functions $P(x,t)$ in Eq.\ (\ref{eqFokker1}) or $P(E,t)$ in Eq.\ (\ref{eqFokker2}) replace the Boltzmann distribution in evaluating the instantaneous probability for electron transfer in the Marcus theory, leading to time-dependent rates. We note that the stationary solution of both Eqs.\ (\ref{eqFokker1}) and (\ref{eqFokker2}) is the Boltzmann distribution, implying that transition state theory will be recovered when relaxation is fast, $\gamma \to 0$ in Eq.\ (\ref{eqFokker1}), or $\gamma\to \infty$ in Eq.\ (\ref{eqFokker2}). 

In what follows, using Eqs.\ (\ref{eqFokker1}) and (\ref{eqFokker2}) as our starting points, we construct numerical simulation procedures for calculating the heat-transport characteristics related to charge transfer processes operating in solvent environments in the corresponding dynamical limits. We investigate their implications for heat deposit in the metal leads and solvent during transfer process. 

\subsection{High-friction regime}
In the high-friction limit ($\gamma \gg \omega_0$) we consider the probability density  $P(x,t|x'_{TR},t_{TR})$ that the reaction coordinate takes the value $x$ following a previous transition event that took place at time $t_{TR}$ at position $x'_{TR}$ of this coordinate. This corresponds to the initial condition $P(x,t_{TR}|x'_{TR},t_{TR})=\delta(x-x'_{TR})$ for which we have found the evolutions (solution of Eq.\ (\ref{eqFokker1})) in the state $A$ and $B$ (Ref. \cite{kir2020})
\begin{align}
\label{eq8} 
&P_A(x,t|x'_{TR},t_{TR})=\sqrt{\frac{D}{2\pi [1-a^2(t-t_{TR})]}} \\ \notag & \times \exp\left\{-\frac{D}{2}\frac{[x-a(t-t_{TR})x'_{TR}]^2}{1-a^2(t-t_{TR})}\right\} \, ,
\end{align}
\begin{align}
\label{eq9}
&P_B(x,t|x'_{TR},t_{TR})=\sqrt{\frac{D}{2\pi  [1-a^2(t-t_{TR})]}} \\ \notag &\times \exp\left\{-\frac{D}{2}\frac{[x-d-a(t-t_{TR})(x'_{TR}-d)]^2}{1-a^2(t-t_{TR})}\right\},
\end{align}
where $D = \beta \hbar \omega_0$ and $a(t)=\exp\left(-\frac{\omega_0^2}{\gamma}t\right)$. For $\gamma\to0$, $a(t)\to 0$ for all time $t>0$, indication "instantaneous" relaxation to an equilibrium Boltzmann distribution in the corresponding wells.

$P_{A}(x,t|x'_{TR},t_{TR})dx$ is the probability to find a solvent configuration with a reaction coordinate in $[x,x+dx]$ for the the unoccupied state $A$ at time $t$, given that the previous transition from the occupied state $B$ has occurred at the solvent configuration $x'_{TR}$ at time $t_{TR}$. Correspondingly, $P_{B}(x,t|x'_{TR},t_{TR})dx$ describes the equivalent for the reduced state $B$. It is important to notice that the next electronic transition can take place at any $x$. This $x$ then becomes the next transition configuration $x_{TR}$ where the electron has energy $\epsilon(x_{TR})$. Consequently the probabilities to find a corresponding metal level occupied $f_K(\epsilon(x_{TR}))$ or unoccupied $1-f_K(\epsilon(x_{TR}))$ are determined from Eq.\ (\ref{eq4}). $f_K(\epsilon)$ is the associated Fermi function 
\begin{align}
f_K(\epsilon)=\frac{1}{\exp\left(\frac{\epsilon+e\Phi_K}{k_BT}\right)+1},
\label{eq10a}
\end{align}
where $\Phi_K$ is the potential of the left($K=L$) or right  ($K=R$) lead, and where $e$, $k_B$ and $T$ are the electron charge, the Boltzmann constant and the temperature, respectively.

Correspondingly, the ET rates (probabilities per unit time), $k_{AB}$ into the molecule, and $k_{BA}$ out of the molecule, are given in this high-friction limit by
\begin{align}
\label{eq6}
k_{AB}^{K}&(t-t_{TR};x'_{TR})= \Gamma \int_{-\infty}^{\infty} dx P_A(x,t|x'_{TR},t_{TR})f_K(x)
\end{align}
\begin{align}
\label{eq7}
k_{BA}^{K}(t-t_{TR};x'_{TR})= \Gamma \int_{-\infty}^{\infty} dx & P_B(x,t|x'_{TR},t_{TR})\\ \notag &\times \left[1-f_K(x)\right]\, .
\end{align}
$\Gamma$ is assumed to be independent of the solvent configuration $x$, while $\Gamma^{-1}$ characterizes the time span between the electronic hopping events. The integration over all solvent configuration $x$ in  Eqs.\ (\ref{eq6}) and (\ref{eq7}) can be extended to $\pm \infty$ because a transition may occur at every solvent configuration along the reaction coordinate, subjected to the Pauli principle that is accounted for explicitly in Eqs.\ (\ref{eq6}) and (\ref{eq7}). 
The limit $\gamma \to 0$ correspond to "infinitely fast" relaxation to equilibrium, the rates given in Eqs.\ (\ref{eq6}) and (\ref{eq7}) become the thermal Marcus rates.

\subsection{Low-friction regime}
In the low-friction regime ($\gamma\ll \omega_0$) we need to solve Eq.\ (\ref{eqFokker2}) to determine solvent relaxation after each electron transfer event.

The probability to find the system at time $t$ with energy $E$ as solution of Eq.\ (\ref{eqFokker2}) where the proceeding transfer happened at time $t_0$ under system energy $E_0$ is given by (Ref.\ \cite{kir2020} with initial $P(E,t_0|E_0,t_0)=\delta(E-E_0)$)
\begin{align}
\label{eq10c}
&P(E,t|E_0,t_0)=\frac{1}{k_B T [1-e^{-\gamma (t-t_0)}]}\times \\ \notag &\exp\bigg[\frac{-[E_0 e^{-\gamma (t-t_0)} +E]}{k_B T [1-e^{-\gamma (t-t_0)}]}\bigg]\sum_{m=0}^\infty \frac{\big[\frac{EE_0e^{-\gamma (t-t_0)}}{k^2_B T^2[1-e^{-\gamma (t-t_0)}]^2}\big ]^m}{m!^2}\Theta(E)\, .
\end{align}

The corresponding rates for the electron insertion and removal process to the respective lead $K$ accompanied by slow energy relaxation finally follow as (see Supplementary Material for details) 
\begin{align}
\label{eq36}
k_{AB}^K&(t-t_0,;E_0)
= \frac{\Gamma}{\sqrt{E_R}} \int_{-\infty}^{\infty} d\epsilon f_K(\epsilon) \\ \notag & \times \int_{0}^{\infty} dE \frac{P(E+\frac{1}{2}\hbar \omega_0 x_{TR}(\epsilon)^2,t|E_0,t_0)}{\sqrt{E}},\\
\label{eq37}
k_{BA}^K&(t-t_0;E_0)
= \frac{\Gamma}{\sqrt{E_R}} \int_{-\infty}^{\infty} d\epsilon (1-f_K(\epsilon)) \\ \notag & \times \int_{0}^{\infty} dE \frac{P(E+\frac{1}{2}\hbar \omega_0 x_{TR}(\epsilon)^2,t|E_0,t_0)}{\sqrt{E}}.
\end{align}
$\Gamma$ is assumed to be independent of the solvent energy. Further $\hbar \omega_0x_{TR}(\epsilon)^2/2$ is energy at the crossing points $x_{TR}(\epsilon)$ of the two energy surfaces which can be read off from Eq.\ (\ref{eq4}). For $\gamma \to \infty$ the rates of Eqs.\ (\ref{eq36}) and (\ref{eq37}) become Marcus' rates.

\section{Numerical method}

We exploit the charge transfer rates of Eqs.\ (\ref{eq6}) and (\ref{eq7}) (high-friction) and Eqs.\ (\ref{eq36}) and (\ref{eq37}) (low-friction) to determine the average heat currents between molecule and the respective lead, as well as between molecule and solvent. To this end we use a numerical Monte Carlo procedure following typical charge transfer trajectories (see numerical Monte Carlo procedure described in Supporting Information in Ref. \cite{kir2020}).

We assume the heat current to be exclusively related to the transferred charges. To quantify the heat currents between the molecule, solvent and leads we consider the explicit heat transfer processes:

(i) $A\to B$: After the electron is transferred from the metal to molecule the respective lead looses the energy $e\Phi_K - \epsilon(x_{TR})$. Note that $\epsilon(x_{TR})$ is defined relative to the Fermi energy $E_F=0$ (in absence of the bias potential $\Phi_K$ the Fermi levels of each electrode $E_F=0$). The solvent configuration for transition $x_{TR}$ or its related electron energy $\epsilon(x_{TR})$ is quantified by Eq.\ (\ref{eq4}) and can be read of from the numerical trajectory for each charge transfer (see Supplementary Material for details). The energy supplied from the solvent at the actual configuration can be read of from the potential surface of Eq.\ (\ref{eq2}) to $-E_A(x_{TR})+E_A+\epsilon(x_{TR})$. Given the previous transition at configuration $x'_{TR}$  the total energy exchange between solvent and molecule is $-E_A(x_{TR})+E_A+\epsilon(x_{TR})+E_B(x'_{TR})-E_B$ which considers the energy already released from solvent $E_B(x'_{TR})-E_B$ (see Eq. (\ref{eq3})). 

(ii) $B\to A$: The electron enters the metal and deposes the energy $ \epsilon(x_{TR})-e\Phi_K$ into the respective lead where it equilibrates very fast in the electronic manifold of the metal. The heat exchange between solvent and molecule reads $-E_B(x_{TR})+E_B+E_A(x'_{TR})-E_A-\epsilon(x'_{TR})$ by exploiting again the free potential surfaces of Eqs.\ (\ref{eq2}) and (\ref{eq3}). 

In the following, we apply a symmetric bias voltage $\Phi_R=-\Phi_L=\Delta \Phi/2$  between the leads with density of states given by Eq.\ (\ref{eq10a}) while the molecular energy level (energy difference between the equilibrium occupied and unoccupiend state) lies $\Delta E$ above the reference Fermi level of both leads when no voltage is applied.
\section{Results}
In this section we present results for the electronic current and the corresponding heat fluxes into the leads and the solvent in molecular junctions in which the electron transport is dominated by the non equilibrium extension of the Marcus electron transfer dynamics as described above. We consider steady state situation in which the electron current in the junction is constant under given voltage bias. In addition to electronic and heat current calculated as functions of the potential bias, we determine the behavior of the thermoelectric Seebeck coefficent for various solvent relaxation dynamics in the high- and low-friction regime.

\subsection{The high-friction limit}

\begin{figure}[h!!!]
\centering
    \begin{subfigure}{\linewidth}
\includegraphics[width=\linewidth]{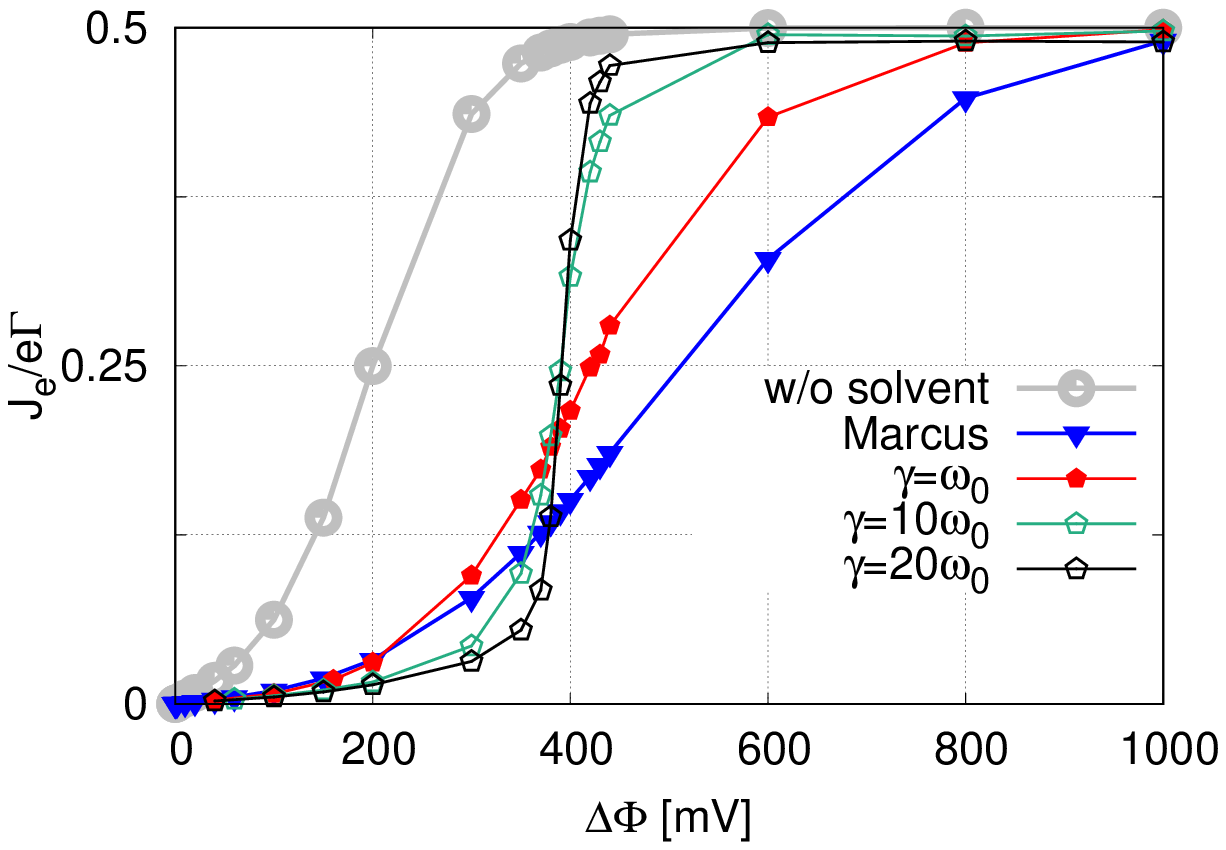} 
    \caption{}
\label{figA}
    \end{subfigure}\hfill
    \begin{subfigure}{\linewidth}
\includegraphics[width=\linewidth]{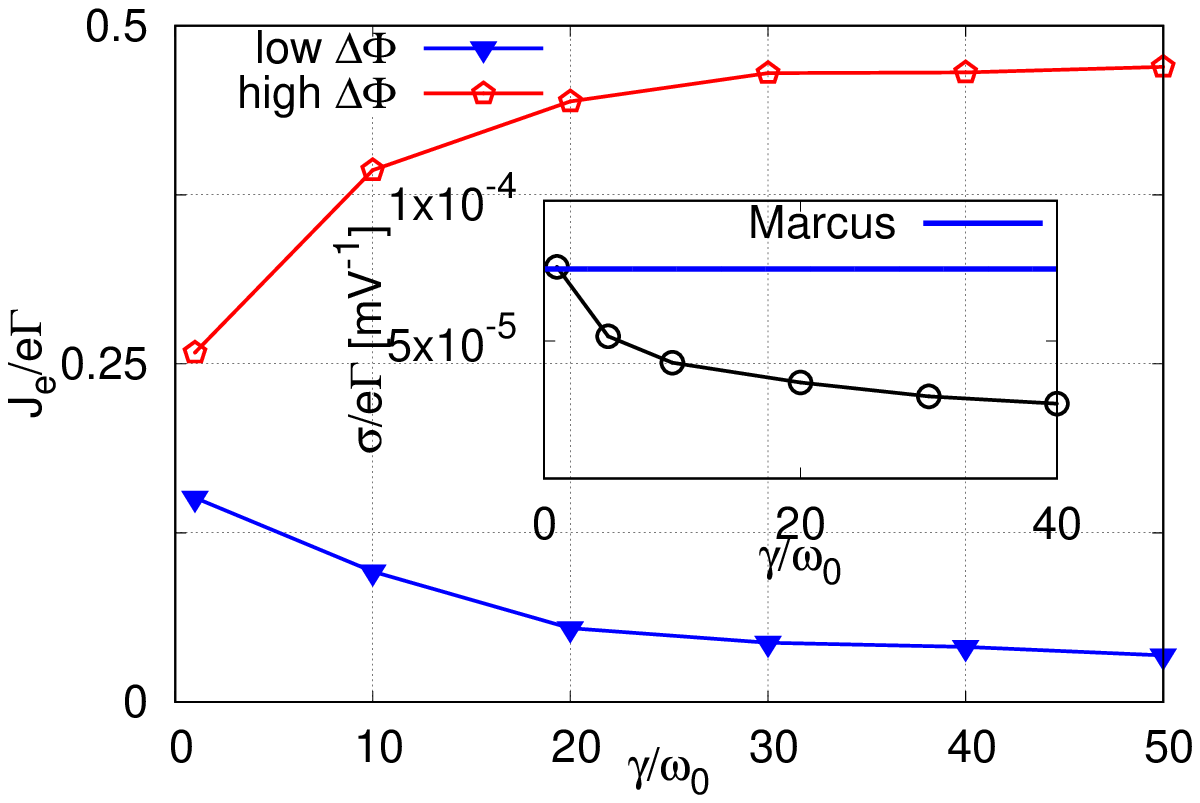}
    \caption{}
\label{figB}
    \end{subfigure}

\caption{(a) Charge current $J_e$ plotted against the bias voltage $\Delta \Phi$ with and without solvent. Solvent dynamics is described using the high-friction model. (b) Charge current $J_e$ plotted against varying friction $\gamma$. $J_e$ is calculated at $\Delta \Phi = 75$mV (labeled as "low $\Delta \Phi$") and $\Delta \Phi = 220$mV (labeled as "high $\Delta \Phi$"). Inset: Conductance $\sigma=dI/d\Delta\Phi|_{\Delta \Phi\to 0}$ plotted against varying friction $\gamma$. The equilibrium energy difference between the molecular "occupied" and "unoccupied" states is $\Delta E = E_B-E_A = 100$meV and the reorganization energy is taken to be $E_R = 200$meV. The temperature T of both leads and solvent is $300 $K. The Marcus regime is reached within the proposed model for a small enough friction, e.g. $\gamma = 0.001 \omega_0$.}
    \end{figure}

Fig.\ (\ref{figA}) illustrates the charge current-voltage characteristics for different choices of the friction $\gamma$ in the high-friction regime, $\gamma \gg \omega_0$. Also shown are the $I$-$V$ behavior in the absence of solvent, as well as in the Marcus limit. As expected, without solvent (Fig.\ (\ref{figA}) grey line) one observes an onset of an appreciable current when the bias potential exceeds the energy difference $\Delta E$ (here taken to $100$meV) between occupied and unoccupied state showing the typical onset of conduction at finite temperature. When immersed in solvent and in the fast relaxation ($\gamma \to 0$, Marcus) limit (Fig.\ (\ref{figA}) blue line) the current-voltage rise at the onset of conduction is less steep than in the absence of solvent as well as in presence of a solvent with $\gamma >0$. This behavior results from the fact that a part of the available electric energy ($e\Delta \Phi$) is needed to overcome the reorganization energy when the molecule changes its charge state during the transfer process (See also discussion in Ref.\cite{mig2011}). In other context this solvent barrier is referred to the Franck-Condon blockade\cite{koc2005,koc2006}

Interestingly, as $\gamma$ increases and solvent relaxation becomes slower (Fig.\ (\ref{figA}) red, green and black lines), the current-voltage step becomes steeper again and reaches its steady value at a threshold voltage (here $\Delta \Phi = 400$mV) smaller than in the Marcus regime. For an extremely high friction, e.g. $\gamma =20\omega_0$, the solvent configuration is nearly frozen and the transition point between molecular state $A$ and $B$ remains almost the same. In order to interchange between the occupied and unoccupied molecular state for sequential charge transfer processes less reorganization energy of the solvent needs to be overcome. The $\gamma \to \infty$ limit is therefore similar to the no-solvent case: once the transition point between the molecular states is accessible by the applied voltage the sequential interchange between both becomes possible. The current then reaches its steady state value on a timescale that is not slowed down by the solvent relaxation dynamics.

Another view of the same data is given by Fig.\ (\ref{figB}) which depicts $J_e$ for varying friction $\gamma$. For small applied voltage, $J_e$ decreases with $\gamma$ (Fig.\ (\ref{figB}) blue curve). This is reminiscent of the high friction limit of Kramers theory for barrier crossing rates \cite{kra1940} and reflects the same physics: the charge transfer rate decreases with increasing friction in this limit. This has been studied in our previous work \cite{kir2020}. In the same way the conductance $\sigma=dI/d\Delta\Phi|_{\Delta \Phi\to 0}$ as charge current per small applied voltage decreases with $\gamma$ (inset in Fig.\ (\ref{figB})). Interestingly, in the high friction limit and for higher applied bias potential, e.g. $\Delta \Phi = 220mV$, $J_e$ increases with $\gamma$ (Fig.\ (\ref{figB}) red curve). This can be understood: With higher friction $\gamma$ the solvent relaxation is reduced and, hence, the more rigid system tends to stay for longer time at the transition configuration along the reaction coordinate. Once the applied bias is high enough, the transition configuration is readily accessible, and so, transition between the molecular states is enhanced which results in $J_e$ increasing with $\gamma$.
\begin{figure}[h!!!]
\centering
    \begin{subfigure}{\linewidth}
\includegraphics[width=\linewidth]{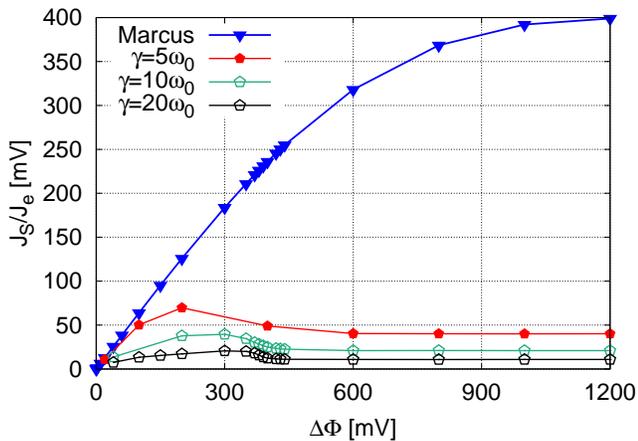} 
    \caption{}
\label{figC}
    \end{subfigure}\hfill
    \begin{subfigure}{\linewidth}
\includegraphics[width=\linewidth]{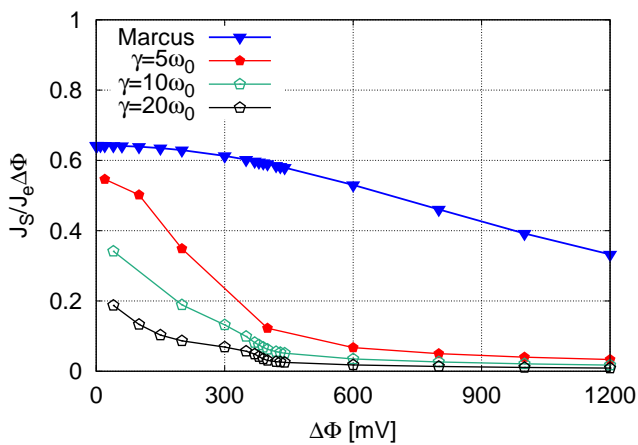}
    \caption{}
\label{figD}
    \end{subfigure}

\caption{(a) The ratio $J_S/J_e$ between the steady state heat flux into the solvent and the electronic current and (b) the ratio between $J_S/(J_e e\Delta \Phi)$ plotted against bias voltage $\Delta \Phi$. Several values of friction using the high-friction model are considered. The equilibrium energy difference between the molecular "occupied" and "unoccupied" states is $\Delta E = E_B - E_A = 100$meV and the reorganization energy $E_R = 200$meV. The temperature $T$ of both leads and solvent is $300$K. The Marcus regime can be obtained within the described model for small enough values of friction, e.g. $ \gamma = 0.001\omega_0$. Note that $J_S > 0$ implies that heat is going into the solvent.}
    \end{figure}

Next, we consider heat fluxes associated with the junction operation. Fig.\ (\ref{figC}) displays the ratio between heat current into solvent and electric current in the junction, essentially, the heat deposited into the solvent per transferred electron. In the Marcus limit this heat flux per transferred electron is seen to increase with the bias potential $\Delta\Phi$ and for high enough voltage, this ratio becomes $2E_R$ (here $= 400$mV, Fig.\ (\ref{figC}) blue line). In this limit, a complete solvent relaxation with energy $E_R$ takes places following each incidence of electron moving between the molecule and an electrode and $2E_R$ per transferred electron just reflect the energy deposited into the solvent in these two steps.

With higher friction $\gamma$ the solvent relaxation is slower and does not adapt completely to the changing molecular charge state during the sequential charge transfer, therefore, $J_S/J_e$ is smaller (Fig.\ (\ref{figC}) red, green and black lines). As a function of $\Delta \Phi$, the ratio first increases linearly in a typical ohmic behavior, then goes through a maximum (for large $\gamma$) before becoming independent of $\Delta \Phi$. This behavior at intermediate values of $\Delta \Phi$ can be explained by noting that in the regime lesser electronic current implies more time for solvent to relax between electron transfer events. For large $\Delta \Phi$ both heat and electronic current, and consequently their ratio, become constant.

The same data is also used in Fig.\ (\ref{figD}) which portrays the ratio $J_S/(J_e\Delta \Phi)$ representing the part of available energy deposited into the solvent. As can be expected, with increasing damping $\gamma$ and reduced solvent relaxation the fraction decreases and larger part of the available energy is deposited into the electrodes. Obviously this partitioning of the available electrical energy also depends on $E_R$. In the context of creating useful electrical motor (e.g in photo-voltaic cells) minimizing the energy flux into the solvent by minimizing the reorganization energy and increasing solvent rigidity will increase the device efficiency \cite{ram2018}.

\begin{figure}[h!!!]
\centering
\includegraphics[width=0.95\linewidth]{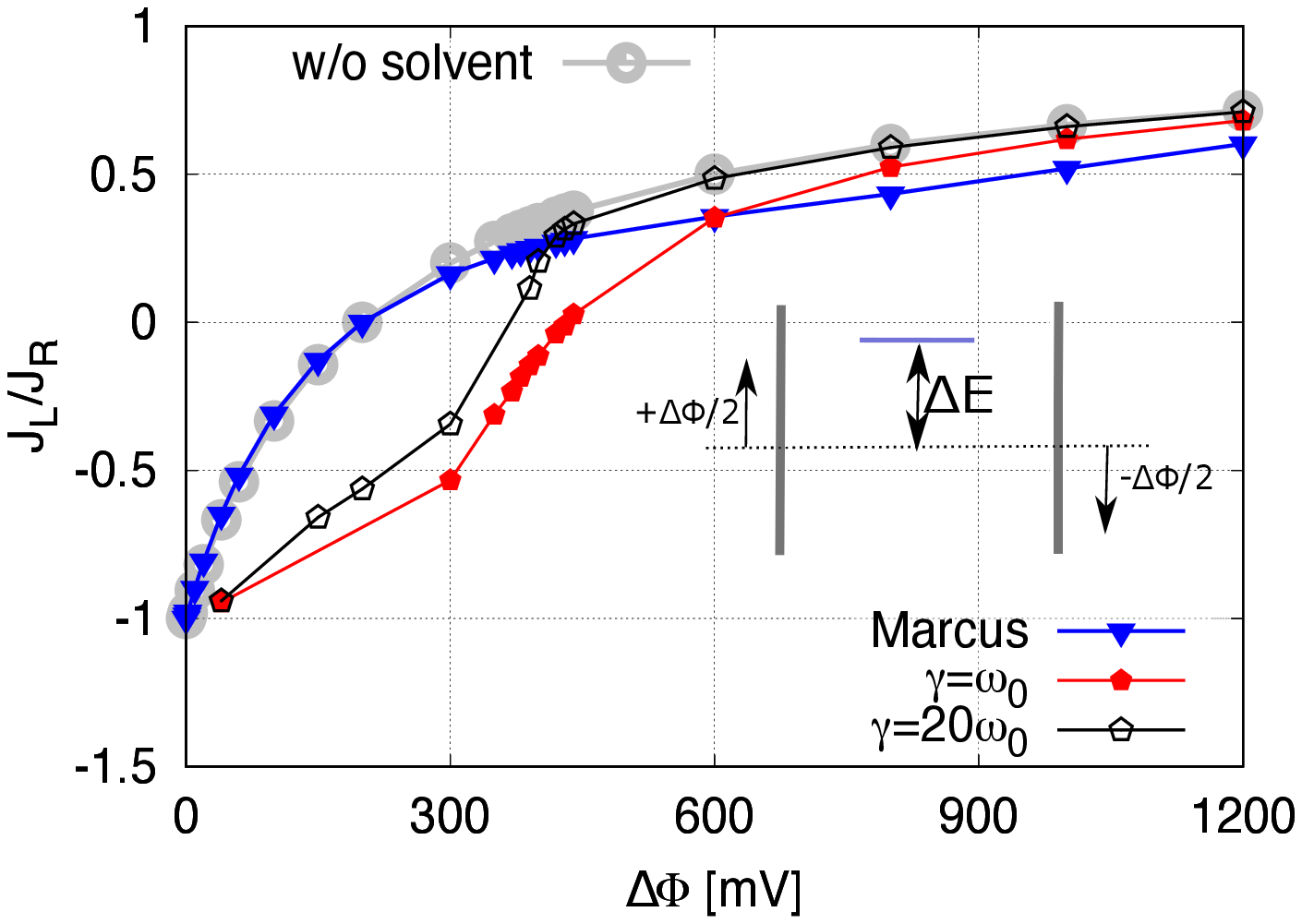} 

\caption{\label{figE}The ratio of $J_L/J_R$ plotted against bias voltage $\Delta \Phi$. The ratio are taken with and without solvent. Within the solvent several values of friction using the high-friction model are considered. The equilibrium energy difference between the molecular "occupied" and "unoccupied" states is $\Delta E = E_B-E_A = 100$meV and the reorganization energy $E_R = 200$meV. The temperature $T$ of both leads and solvent is $300$K. The Marcus regime can be obtained within the described model for small enough values of friction $\gamma \to 0$. Inset: Molecular junction with $\Delta E$ above the equilibrium chemical potential of the leads when no voltage is applied.}
\end{figure}

The heat added to or removed from a given electrode is a combination of the heat converted from the available electrical energy $e\Delta \Phi$ and the heat that is transferred between the leads. For the specific example of molecular junction shown in the inset, Fig.\ (\ref{figE}) depicts the ratio of heat deposited into the left and right electrodes for different applied bias voltages \cite{note}. In the limit $\Delta \Phi \to 0$ the ration approaches $-1$ since in this limit heat transfer between the electrodes is the dominating process. In addition, the part of the electrical energy converted into heat, $e(\Delta \Phi-J_S /J_e)$, is also distributed between the electrodes so that both $J_L$ and $J_R$ and their ratio become positive for large $\Delta \Phi$. With increasing solvent damping $\gamma$ (Fig.\ (\ref{figE}) red/black vs. blue/grey curves) more voltage needs to be applied to make this ratio positive.


\subsection{The low-friction limit}
In the low-friction limit, $\gamma \ll \omega_0$, the characteristic time for energy loss by the reaction coordinate is slow relative to its internal dynamics \cite{NitzanBook,kra1940}. We therefore utilize this energy E as a dynamical variable whose probability distribution evolves according to Eq.\ (\ref{eq10c}) and in turn determines the charge transfer rates according to Eqs.\ (\ref{eq36}) and (\ref{eq37}).

\begin{figure}[h!!!]
\centering
    \begin{subfigure}{\linewidth}
\includegraphics[width=\linewidth]{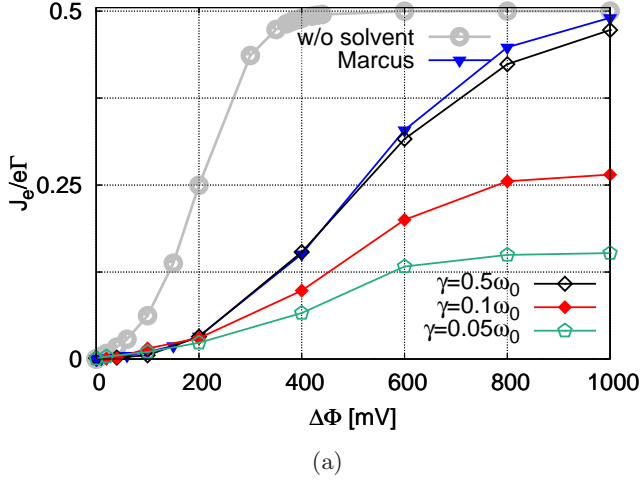} 
    \caption{}
\label{figF}
    \end{subfigure}\hfill
    \begin{subfigure}{\linewidth}
\includegraphics[width=\linewidth]{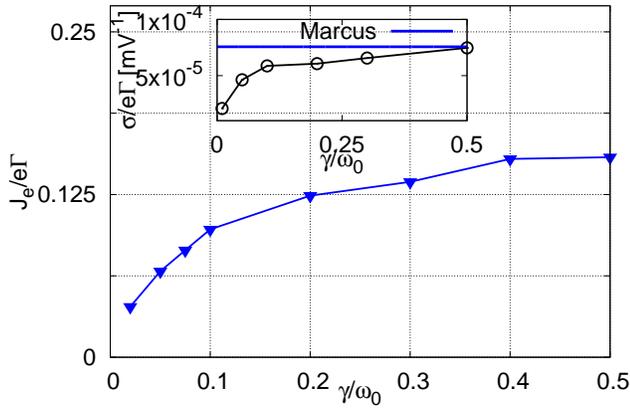}
    \caption{}
\label{figG}
    \end{subfigure}

\caption{(a) Charge current $J_e$ plotted against the bias voltage $\Delta \Phi$ with and without solvent. Solvent dynamics is described using the low-friction model. (b) $J_e$ plotted against varying friction $\gamma$. $J_e$ is calculated at $\Delta \Phi = 200$mV. Inset: Conductance $\sigma=dI/d\Delta\Phi|_{\Delta \Phi\to 0}$ plotted against varying friction $\gamma$. The equilibrium energy difference between the molecular "occupied" and "unoccupied" states is $\Delta E = E_B-E_A = 100$meV and the reorganization energy is taken to be $E_R = 200$meV. The temperature T of both leads and solvent is $300 $K. In this model, the Marcus regime is approached in the large $\gamma$ limit, $\gamma > \omega_0$.}
    \end{figure}

Fig.\ (\ref{figF}), the analog of Fig.\ (\ref{figA}), depicts the current-voltage behavior computed for this regime. As expected, the electric current increases with higher applied voltage and is maximized in the Marcus (large $\gamma$, fast relaxation) limit. The current $J_e$ decreases when we move in the direction of smaller $\gamma$ in analogy to Kramers barrier crossing \cite{kra1940} which is depicted in Fig.\ (\ref{figG}). In the same way the conductance $\sigma$ decreases with smaller $\gamma$ (inset Fig.\ (\ref{figG})). 

\begin{figure}[h!!!]
\centering
    \begin{subfigure}{\linewidth}
\includegraphics[width=\linewidth]{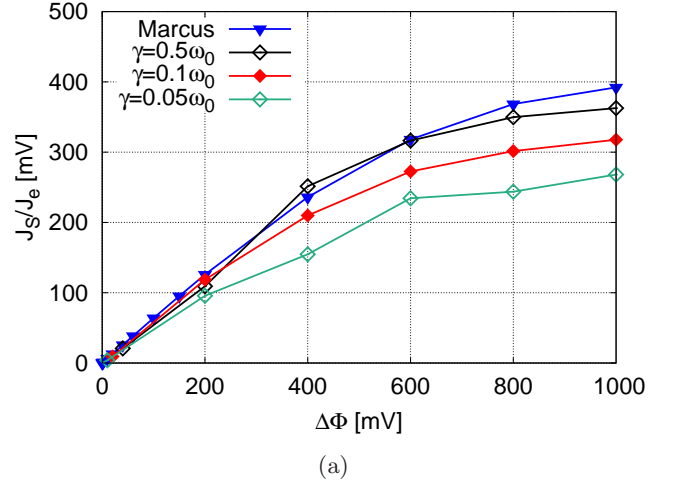} 
    \caption{}
\label{figH}
    \end{subfigure}\hfill
    \begin{subfigure}{\linewidth}
\includegraphics[width=\linewidth]{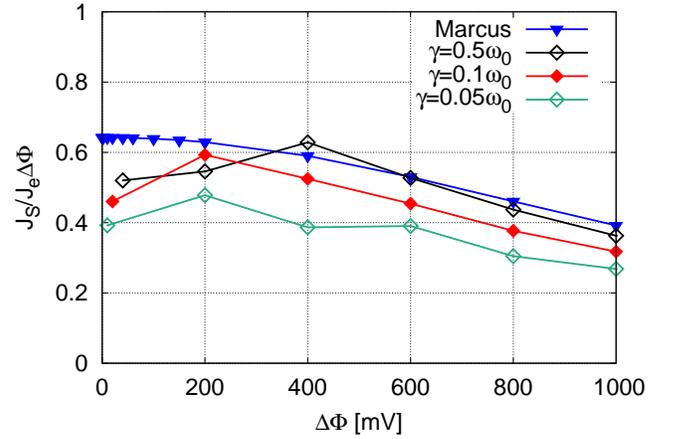}
    \caption{}
\label{figI}
    \end{subfigure}

\caption{(a) The ratio $J_S/J_e$ between the steady state heat flux into the solvent and the electronic current and (b) the ratio between $J_S/(J_e \Delta \Phi)$ plotted against bias voltage $\Delta \Phi$. Several values of friction using the low-friction model are considered. The equilibrium energy difference between the molecular "occupied" and "unoccupied" states is $\Delta E = E_B - E_A = 100$meV and the reorganization energy $E_R = 200$meV. The temperature $T$ of both leads and solvent is $300$K. Note that $J_S > 0$ implies that heat is going into the solvent. In this model, the Marcus regime is approached in the large $\gamma$ limit, $\gamma > \omega_0$.}
    \end{figure}

Fig.\ (5), the analog of Fig.\ (2) for the high friction limit, shows two views of $J_S$ - the heat current deposited into the solvent. Fig.\ (\ref{figH}) depicts $J_S$ per transferred electron while Fig.\ (\ref{figI})  shows it as the fraction of the total available energy $e\Delta\Phi$. As expected, when $\gamma$ is small, the energy relaxation of the reaction coordinate is reduced, and so, the heat transfer into the solvent. Consequently, the part of the available electrical energy $e\Delta\Phi$ going into the solvent is reduced for small $\gamma$ as seen in Fig.\ (\ref{figI}) .

\begin{figure}[h!!!]
\centering
\includegraphics[width=\linewidth]{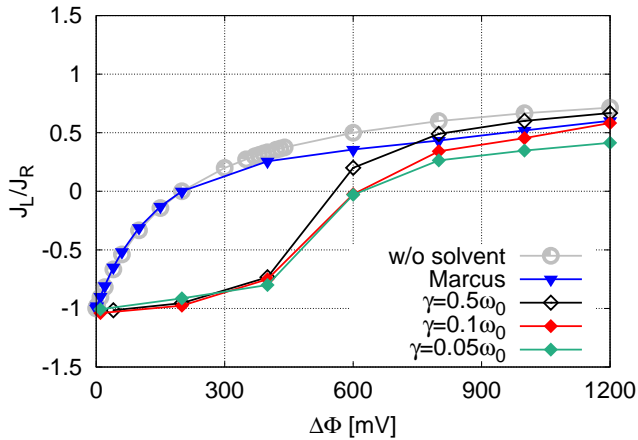} 

\caption{\label{figJ} The ratio of $J_L/J_R$ plotted against bias voltage $\Delta \Phi$. The ratio are taken with and without solvent. Within the solvent several values of friction using the low-friction model are considered. The equilibrium energy difference between the molecular "occupied" and "unoccupied" states is $\Delta E = E_B-E_A = 100$meV and the reorganization energy $E_R = 200$meV. The temperature $T$ of both leads and solvent is $300$K. In this model, the Marcus regime is approached in the large $\gamma$ limit, $\gamma > \omega_0$.}
    \end{figure}
    
The ratio $J_L/J_R$ between the heat currents deposited into the left and right lead is depicted in Fig.\ (\ref{figJ})  (analog of Fig.\ (\ref{figE})) for the molecular junction parameters. As before, the small voltage regime, the heat deposited into the left and right lead is dominated by heat transfer between both leads, so one lead is cooled while the other is heated, resulting in negative $J_L/J_R$ \cite{note}. With increasing $\Delta \Phi$ the heat generated by the electric energy becomes dominant leading to net heating of both electrodes which makes $J_L/J_R$ positive. For lower damping $\gamma$, a higher voltage $\Delta \Phi$ needs to be applied to obtain a positive ratio $J_L/J_R$, such that heat transfer between both electrodes remains the dominant process in a larger regime of $\Delta \Phi$.


\section{Application to calorimetric device}

We now investigate the power input into a prototype calorimeter based on a the metal-molecule-metal junction immersed in the solvent within the high- and low friction regime.
We utilize the calorimeter model of Cui et al.\cite{cui2018}: In that work, an atomic force microscope (AFM) tip was used as a probe electrode to measure the heat deposited into a counter-(calorimeter) electrode - a gold coated calorimetric microdevice. Here, we study the heating or cooling power deposited in one of the electrodes of our specific junction (inset in Fig.\ (\ref{figE})).
\begin{figure}[h!!!]
\centering
\includegraphics[width=\linewidth]{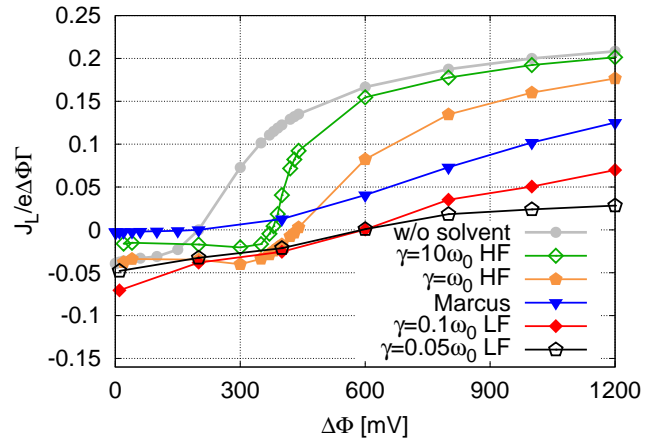} 

\caption{\label{figK} Cooling and heating power of the left electrode $J_L$ per available electric energy $e \Delta \Phi$ for several imposed frictions using the high-friction (HF) and low-friction (LF) model and within the Marcus regime. The equilibrium energy difference between the states is $ \Delta E = E_B-E_A = 100$meV and the reorganization energy were chosen to $E_R = 200$meV. The temperature $T$ of both leads and solvent is $300$K.}
    \end{figure}
Specifically, we choose the left electrode ($K = L$) as calorimeter electrode and study the corresponding heating/cooling power calculated for this electrode.
It is important to realize (see Ref.\cite{cui2018}) that the heat generated on either electrode has a component that purely reflects heat transfer between the electrodes, and one that arises from energy conversion ($e \Delta \Phi$ shared between the electrodes). Writing $J_L$ as a power series in $\Delta \Phi$, $J_L=J_L'\Delta \Phi + J_L'' \Delta \Phi^2+\cdots$, the linear arises from energy transfer and is given by $J_L'=ST\sigma \Delta \Phi$ where $S$ is the Seebeck coefficient, $T$ is the temperature and $\sigma$ is the electrical conduction (at zero bias) \cite{cui2018}. The second order term, $J_L''$, is related to the generated heat. In Fig.\ (\ref{figK}) we show $J_L/e\Delta \Phi$ as a function of $\Delta \Phi$. The intersection of these lines with the $\Delta \Phi = 0$ axis reflect the linear contribution, proportional to the Seebeck coefficient, conductivity and temperature, while the slope corresponds to the electric energy conversion to heat. For small enough $\Delta \Phi$ energy transfer dominates and, in the setup used in  Fig.\ (\ref{figK}) (shown in the inset to Fig.\ (\ref{figE})), the calorimetric electrode shows a cooling behavior $J_L < 0$ for $\Delta \Phi > 0$. Beyond a certain voltage, heat generation dominates and both electrodes undergo heating, so $J_L > 0$. Quantitative aspects of this behavior can be tuned by the solvent friction. With increasing $\gamma$ in the high friction limit and decreasing $\gamma$ in low friction limit, cooling power is increased in comparison to the Marcus regime. Obviously, energy conversion of the available electric energy current $J_e\Delta\Phi$ into heat is divided between the electrodes and the solvent as already described in the sections above. Interestingly, also the energy transfer between electrodes shows a strong dependence on $\gamma$.
\section{Seebeck coefficient}
Next, we consider the thermoelectric behavior of the molecular junction as measured by the Seebeck coefficient \cite{ben1992, sow2019}
\begin{align}
\label{eqSee}
S=-\left.\lim_{\Delta T \to 0} \frac{V_{th}}{\Delta T}\right\vert_{J_e=0},
\end{align}
where the thermal voltage $V_{th}$ is the open circuit voltage induced in the junction by applying a temperature difference $\Delta T$ between the two electrodes. In the calculation reported below we set the temperature of one electrode and the solvent to $T$ and the temperature of the other lead to $T+\Delta T$. $V_{th}$ is calculated as potential difference $\Delta \Phi$ which nullifies the charge current for a given $\Delta T$, and the limit (\ref{eqSee}) is taken to obtain $S$.
\begin{figure}[h!!!]
\centering
    \begin{subfigure}{\linewidth}
\includegraphics[width=\linewidth]{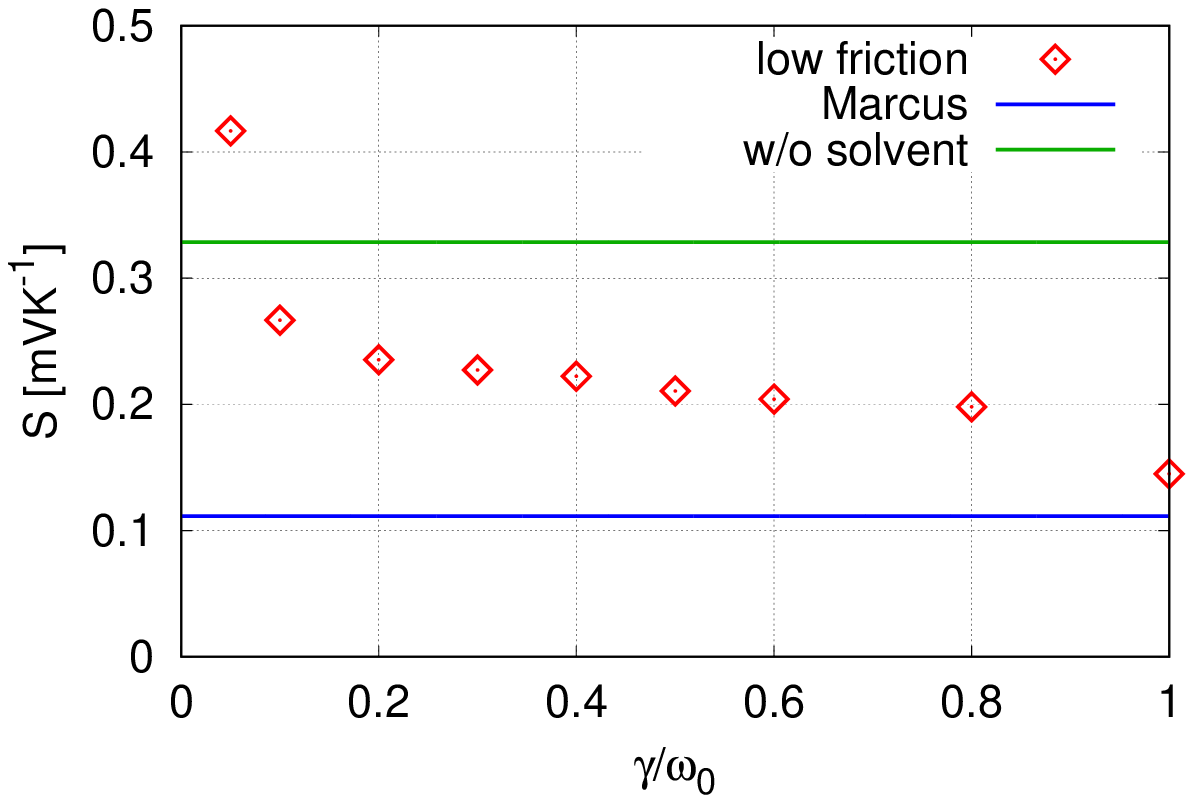} 
    \caption{}
\label{figL}
    \end{subfigure}\hfill
    \begin{subfigure}{\linewidth}
\includegraphics[width=\linewidth]{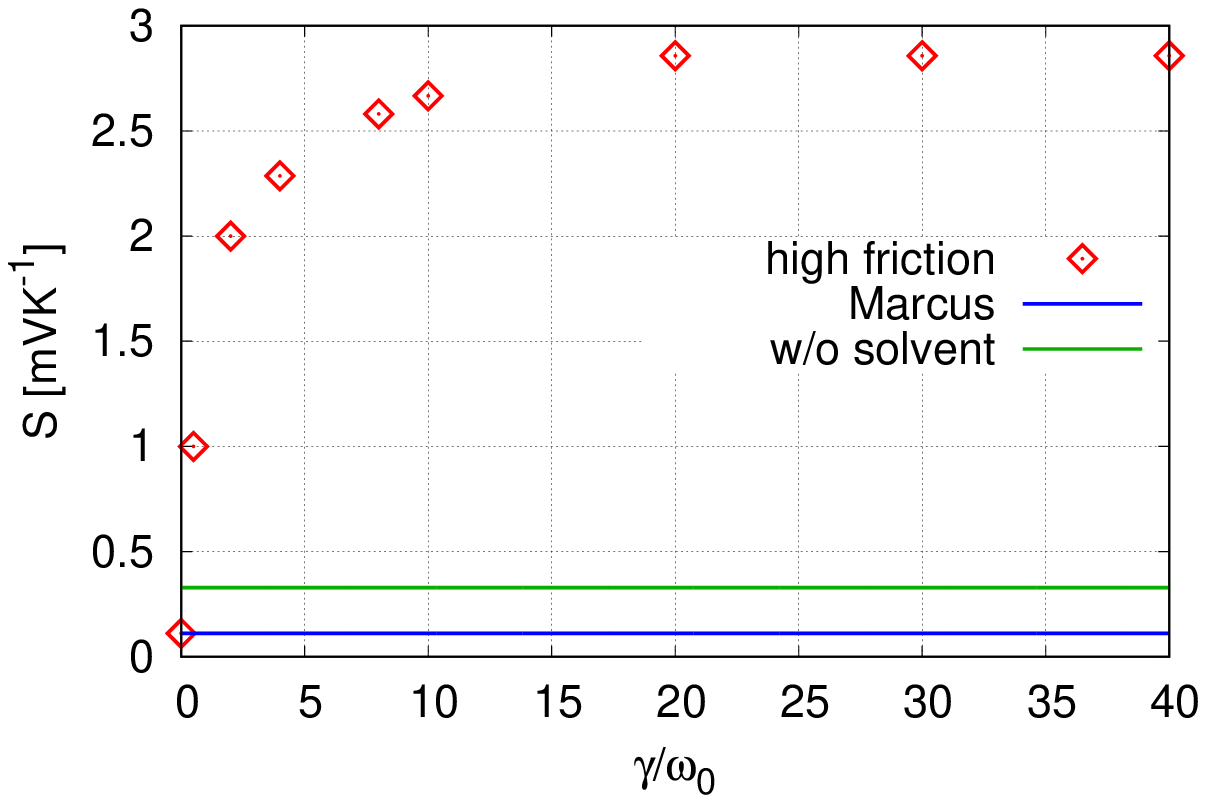}
    \caption{}
\label{figM}
    \end{subfigure}

\caption{Seebeck coefficient in the low friction and (b) high friction model. The blue and green line indicates the Seebeck coefficient for the Marcus model and model without solvent respectively. The equilibrium energy difference between the states is $\Delta E= E_B-E_A=100$meV and the reorganization energy $E_R=200$meV. The temperature $T$ of the right leads and solvent is $300$K, while we slightly change the temperature of the left lead by $T+\Delta T$ to determine the potential difference $\Delta \Phi$ which nullifies the current for a given $\Delta T$. The resulting ratio $\Delta\Phi/\Delta T$ defines the value of the Seebeck coefficient $S$.}
    \end{figure}

The results of this calculation are shown in Figs.\ (\ref{figL}) and (\ref{figM}), which  portray the dependence of Seebeck coefficient on solvent friction $\gamma$ in both the high- and low-friction regimes.

A vanishing friction in the low friction limit and an increased friction in the high friction limit enhance the Seebeck coefficient. This require a higher applied bias $\Delta \Phi$ to nullify the charge current induced by a given temperature difference $\Delta T$ across the junction in comparison to the Marcus regime (recall that the Marcus limit is given by $\gamma \geq 1$ for low-friction and $\gamma \ll 1$ for the high-friction regime). The following observations are noteworthy:

(a) Although the Seebeck coefficient is a zero current property of the system it depends on the solvent friction $\gamma$ that could be expected not to be expressed in the zero current behavior of the system. We should recall however that zero charge current here is not an expression of equilibrium but of balancing  out of equilibrium of two different drivings (temperature difference $\Delta T$ and solvent friction $\gamma$). A similar behavior is found in models of photovoltaic cells where the open circuit photo voltage is found to depend on the relaxation rates associated with the ambient and the sun temperatures \cite{kir2022}, and is analogous to experimental and theoretical observations of the Seebeck coefficient dependence on electron-phonon interaction \cite{gal2008,kim2014,zim2016}.

(b) The Seebeck coefficient increases when moving from the Marcus limit into the high- and low-friction regimes. It appears that, although the overall net charge current vanishes, charge migration in the two directions through the molecule at at range of configurations that is determined by the solvent as it adapts to the molecular charging state. The reduced solvent relaxation between charge migration processes - directly associated with $\gamma$ - restricts the possible transition configurations (values of the coordinate $x$ or the energy $E$ of the reaction coordinate in the high- and low-friction limits, respectively). Only with a bias $\Delta \Phi$ higher than in Marcus regime (full solvent relaxation), one can reach the restricted transition configuration where voltage and thermal driving of net current can balance out.

(c) It is interesting to note that the corresponding Seebeck coefficient in high- and low-friction regime can exceed its value of the junction without solvent. In the low-friction regime this is seen to happen in the limit $\gamma \to 0$. Note that a vanishing $\gamma$ does not mean an absent solvent but a small damping of the energy $E$ within the potential associated with the "occupied" or "unoccupied" molecular states (see Eq.\ (\ref{eq10c})). The continuing configuration interchange between the two states is affected by $\gamma$ and is expressed by the $\gamma$-dependence in the way the charge current $J_e$ is affected by $\Delta \Phi$ (see conductance in inset of Fig.\ (\ref{figG})) and $\Delta T$. The corresponding effect on $S=\Delta\Phi/\Delta T|_{\Delta T \to 0}$ is seen in Fig.\ (\ref{figL}). The same arguments as above can be used when considering the high-friction limit where a larger $\gamma$ "freezes" more the transition configuration, in this limit, described by the solvent coordinate $x$. The Seebeck coefficient increases with $\gamma$ and exceed its value junction without solvent (Fig.\ (\ref{figM})).

\section{Conclusion}
We have shown that in "underwater" molecular junctions, in the limit where electron transport is dominated by Marcus-type electron transfer steps, that solvent dynamics, which in our model is expressed by a solvent friction, can strongly affect the electrical, thermal and thermoelectric junction properties. A reduced heat dissipation into the solvent ensues a higher heat transfer between the leads which also depends on the given junction setup and the reorganization energy of the solvent. This imply that the junction transport behavior can be controlled by solvent properties. In particular, we have suggested in our previous work \cite{kir2020} that $\gamma$ may depend on the solvent dielectric properties. These can be tuned by confining the solvent down several nanometers, e.g., by decreasing the height of water in between to glassy slaps \cite{fum2018}. New (experimental) realizations might be possible by utilizing the solvent friction $\gamma$ to built efficient molecular-based thermoengines. 
We have further shown that the Seebeck coefficient of a prototype molecular junction grows significantly with varying $\gamma$ in the limiting regimes of high-(slow solvent motion) and low-(underdamped solvent motion) friction. This observation is reminiscent of previous works that show dependence of the Seebeck coefficient on electron-phonon interaction \cite{gal2008,zim2016}. Observing such behavior in "underwater" junctions in the presence of solvent friction suggests the use of the latter as a new tool to control the Seebeck coefficient of immersed molecular junction. Further specifications of the model, e.g., the inclusion of lifetime broadening as shown in Refs.\cite{sow2018,sow2019} will be addressed in future work.
In total, the consideration of solvent dynamics may be not only a promising way to improve the setup of thermo-nanodevices based on single molecular junctions, but also should stimulate further systematic exploration of atomic- and molecular-scale thermal transport.

\section*{Supplementary Material}
See Supplementary Material for detailed derivation of Eqs.\ (\ref{eq36}) and (\ref{eq37}) and the protocol to register heat transfer in our numerical simulation.

\section*{Acknowledgments}
This work has been supported the U.S. National Science Foundation under the Grant No. CHE1953701 and the University of Pennsylvania. 

\section*{Data Availability}
The data that support the findings of this study are available from the corresponding author upon reasonable request.


\bibliography{ET}

\begin{thebibliography}{39}
\bibitem [{not()}]{note}%
  \BibitemOpen
  \href@noop {} {\bibinfo  {journal} {Note that this ratio depends on the way
  the potential falls between the two leads as well as the position of the
  molecular level relative to the equilibrium chemical potential of the lead}\
  }\BibitemShut {NoStop}%
\end{thebibliography}

\clearpage 

\onecolumngrid
\appendix

\section*{Supplementary Material on \\Energy Transfer and Thermoelectricity in Molecular Junctions in Non-Equilibrated Solvents}

In this Supplementary Material we derive in the first section the charge transfer rates in the low-friction limit (Eqs.\ (14) and (15)). In the second section we present an exemplary protocol exploited to determine the heat current (heat exchange per time) in our simulation. 

\section{Transition rates in low-friction regime}
\label{TRlowfr}

For a charge transfer where the solvent imposes a low-friction ($\gamma\ll \omega_0$), we have found the energy probability distribution
\begin{align}
P(E,t|E_0,t_0=0)&=h\sum_{n=0}^\infty \frac{(-h)^n[E_0e^{-\gamma t}+E]^n}{n!}\cdot \sum_{m=0}^{\infty}\frac{h^{2m}E^m E_0^m e^{-m\gamma t}}{m!^2} \\
&=\frac{1}{k_B T [1-e^{-\gamma t}]}\exp\bigg[\frac{-[E_0 e^{-\gamma t} +E]}{k_B T [1-e^{-\gamma t}]}\bigg]\cdot \sum_{m=0}^\infty \frac{\big[\frac{EE_0e^{-\gamma t}}{k^2_B T^2[1-e^{-\gamma t}]^2}\big ]^m}{m!^2},
\end{align}
as solution of the Smoluchowski-like equation for an energy $E$ at time time $t$ given that we have a transition energy $E_0$ at $t_0$ (see details of derivation in our previous work \cite{kir2020}). 

One way to determine the charge transmission rate, e.g. from state $A\to B$ (occupation),  can be the following (see Ref.\ \cite{NitzanBook} , Eq.\ (16.5)):

\begin{align}
\label{rate}
k_{A\to B}= \int_0^\infty d \dot{x} \dot{x} P(x_{TR},\dot{x}) P_{A\to B}(\dot{x}),
\end{align}
where $\dot{x}$ is the velocity to move in the potential surface of state $A$ (in one direction) and $x_{TR}$ is the transition point to the potential surface of state $B$. $P(x_{TR},\dot{x})$ is the probability density to be at the transition point $x_{TR}$ with velocity $\dot{x}$.  $P_{A\to B}(\dot{x})$ is the transition probability to go from one surface to the other of Landau-Zener type (Eq.\ (16.6) in Ref.\ \citep{NitzanBook}). The potential surfaces are characterized by the Marcus parabola $E_A(x, \epsilon)=\frac{1}{2}\hbar \omega_0 x^2 +E_A+\epsilon$ (Eq.\ (1) in main text) and $E_B(x)=\frac{1}{2}\hbar (x-d)^2+E_B$ (Eq.\ (2) in main text), where $\epsilon$ is the energy of the transferred charge and $d$ is the respective shift of both parabolas. Since we are in the non-adiabatic regime (small molecule-lead coupling), energy conversation requires the transition at the parabolas crossing
\begin{align}
\label{crossing}
x_{TR}=(E_B-E_A-\epsilon+E_R)/\hbar\omega_0d
\end{align}
with the reorganization energy $E_R=\frac{1}{2}\omega_0 d^2$. 

In the non-adiabatic regime the transition from $A$ to $B$ is rare, so we approximate the transition probability $P_{A\to B}=\frac{2\pi |V|^2}{\hbar |\dot{x}\Delta F|}$ with the interstate coupling $V$ \cite{lan1932,zen1932}. We determine the difference in surface slopes at transition point as $\Delta F= \partial/\partial x (E_A(x)-E_B(x))_{x_{TR}}=\hbar \omega_0 d = \sqrt{h\omega_0} \sqrt{2E_R}$.

The total solvent energy can be split in its kinetic and potential energy part as 
\begin{align}
\label{energy}
E=\frac{1}{2}\hbar \omega_0^{-1} \dot{x}^2 +\frac{1}{2} \hbar \omega_0 x_{TR}^2,
\end{align}
where $x$ is dimensionless counterpart of $\tilde{x}=\sqrt{\frac{\hbar}{m\omega_0}}x$. 

So
\begin{align}
\label{derivative}
dE=\hbar \omega_0^{-1} \dot{x}d\dot{x}
\end{align}
and
\begin{align}
\label{velocity}
\dot{x}=\sqrt{2\omega_0\hbar^{-1}(E-\frac{1}{2}\hbar x_{TR}^2)}.
\end{align}
Furthermore, we set $P(\dot{x},x)\equiv \hbar P(E)/\pi$ which conserve units and normalization of the probability density. 

[See normalization check: 

Suppose $P(\dot{x},x_{TR})=e^{-\frac{1}{2}\hbar \omega_0^{-1} \dot{x}^2-\frac{1}{2} \hbar \omega_0 x_{TR}^2}$ and $P(E)=e^{-E}\equiv e^{-\frac{1}{2}\hbar \omega_0^{-1} \dot{x}^2-\frac{1}{2} \hbar \omega_0 x_{TR}^2}$. When calculating the integral on both sides over the respective degrees of freedom one finds the relation $\int_{-\infty}^{\infty} dX e^{-X^2} \int_{0}^{\infty} d\dot{X} e^{-\dot{X}^2} \equiv \pi \hbar^{-1} \int_0^\infty dE e^{-E}$ with $X^2=\frac{1}{2} \hbar \omega_0 x_{TR}^2$ and $\dot{X}^2=\frac{1}{2}\hbar \omega_0^{-1} \dot{x}^2$.]

Then, the rate in Eq.\ (\ref{rate}) can be written as
\begin{align}
\label{rate2}
k_{A \to B} = \frac{|V|^2}{\hbar \sqrt{E_R}} \int_{\frac{1}{2}\hbar \omega_0^2 x_{TR}^2}^{\infty} dE \frac{P(E,t|E_0)}{\sqrt{E-\frac{1}{2}\hbar \omega_0 x_{TR}^2}} = \frac{|V|^2}{\hbar \sqrt{E_R}} \int_{0}^{\infty} dE \frac{P(E+\frac{1}{2}\hbar \omega_0 x_{TR}^2,t|E_0)}{\sqrt{E}}.
\end{align}
Now we sum over all electrons with individual energy $\epsilon$ from the respective lead $K=L,R$, related to $x_{TR}(\epsilon)$ (Eq.\ (\ref{crossing})), where the electron energy obeys Fermi distribution $f_K(\epsilon)$. Thus, the rate in Eq.\ (\ref{rate2}) reads 
\begin{align}
\label{rate3}
k_{A \to B}^K = \frac{ |V|^2}{\hbar \sqrt{E_R}} \int_{-\infty}^{\infty} d\epsilon \rho f_K(\epsilon) \int_{0}^{\infty} dE \frac{P(E+\frac{1}{2}\hbar \omega_0 x_{TR}(\epsilon)^2,t|E_0)}{\sqrt{E}}. 
\end{align}
After performing $\sum f(\epsilon) = \int d\epsilon \rho f(\epsilon)$, where $\rho$ is the electron density of states, assumed to be constant, we define the rate of electron transfer rate as $\Gamma=|V|^2\rho/\hbar$. Exploiting this definition we obtain Eq.\ (14) in the main text.

We confirm for infinitely fast solvent energy relaxation, $\gamma\to \infty$,
\begin{align}
k_{A \to B} = \Gamma \sqrt{\frac{\pi}{E_Rk_BT}} \int_{-\infty}^{\infty} d\epsilon f(\epsilon) e^{-\frac{(E_B-E_A+E_R-\epsilon)^2}{4k_BTE_R}},
\end{align}
the conventional Marcus transfer rate.

With the same arguments as above we can determine the transfer rate of Eq.\ (15) in the main text to
\begin{align}
\label{rate3}
k_{B \to A}^K = \frac{\Gamma}{\sqrt{E_R}} \int_{-\infty}^{\infty} d\epsilon (1-f_K(\epsilon)) \int_{0}^{\infty} dE \frac{P(E+\frac{1}{2}\hbar \omega_0 x_{TR}(\epsilon)^2,t|E_0)}{\sqrt{E}}.
\end{align}

\newpage
\section{Numerical protocol on energy transfer}
\label{NP}
The following table illustrates exemplarily the energy transfer between molecule, solvent and leads when the molecule becomes occupied $A\to B$ from left lead \textcolor{blue}{[alternatively from right lead]} at time $t_i$ while the previous deoccupation $B\to A$ has happened at time $t_{i-1}$ to right \textcolor{blue}{[alternatively to left]} lead:

\begin{center}
\begin{tabular}{ |p{4cm}||p{5.8cm}|p{5.8cm}| }
 \hline
 \multicolumn{3}{|c|}{Energy transfer} \\
 \hline
  Process & $B\to A$ (deoccupation) & $A \to B$ (occupation) \\
 \hline
 Time ($t_i\geq t_{i-1}$) & $t_{i-1}$ & $t_{i}$ \\
 \hline
 Transition point (corresponding electron energy) (Eq.\ (4) in main text)& $x_{TR}(\epsilon(t_{i-1}))$ & $x_{TR}(\epsilon(t_{i}))$\\
 \hline
 Solvent energy (see Eq.\ (1) in main text)   & $E_S(t_{i-1})=E_B(x_{TR}(\epsilon(t_{i-1}))-E_B+\epsilon(t_{i-1})$&   $E_S(t_{i})=E_A(x_{TR}(\epsilon(t_{i}))-E_A-\epsilon(t_{i})$\\
  \hline
(i) Solvent energy per time &  \multicolumn{2}{|c|}{$\Delta E_S=E_S(t_{i})-E_S(t_{i-1})/(t_i-t_{i-1})$}\\
 \hline
(ii) Energy exchange with left lead & \textcolor{blue}{[$\epsilon_L=\epsilon(t_{i-1})-\mu_L$]} & $\epsilon_L=-\epsilon(t_{i})+\mu_L$\\
  \hline
Energy exchange with right lead & $\epsilon_R=\epsilon(t_{i-1})-\mu_R$  & \textcolor{blue}{[$\epsilon_R=-\epsilon(t_{i})+\mu_R$]} \\
  \hline
\end{tabular}
\end{center}

Now we are able to determine the average energy exchange per time or heat current in steady state. The resulting average heat current into the solvent is obtained by taking the average of $\langle \Delta E_{S}\rangle = \sum_i \Delta E_{S,i}/N$ (line (i)) where $N$ is the total number of occupation and deoccupation events.

In order to obtain the average heat current with the left lead, we first sum over all energies $\epsilon_{L}$ of all events $N_L$ with respect to the left lead and calculate its average according to $\langle \epsilon_L\rangle = \sum_k \epsilon_{L,k} /N_L $ (line (ii)). The average energy divided by the average time $T_L$ elapsed since a previous event with respect to the left lead $\langle \epsilon_L\rangle/T_L$ gives the heat current to the left lead. In steady state and by arguments of ergodicity we can separately perform average over energy and time before determine the resulting average heat current.  Note that the molecule - although less probably with higher applied bias - can also be deoccupied at $t_{i-1}$ from the left lead, illustrated in blue. 

The same calculation is performed to obtain the average heat current with the right lead $\langle \epsilon_R\rangle/T_R$.




%
%
%
%
%

\end{document}